\documentclass[prl,amsmath,aps,twocolumn,superscriptaddress]{revtex4}
\usepackage{graphicx,xspace}
\usepackage{float}
\usepackage{amsmath}
\usepackage{amssymb}
\usepackage[normalem]{ulem}
\input{epsf}
\begin{document}
\newcommand{\scalar}[2]{\left \langle#1\ #2\right \rangle}
\newcommand{\me}{\mathrm{e}}
\newcommand{\mi}{\mathrm{i}}
\newcommand{\dif}{\mathrm{d}}
\newcommand{\period}{\text{per}}
\newcommand{\free}{\text{fr}}
\newcommand{\eq}[1]{Eq.~(\ref{e:#1})}
\newcommand{\eqq}[1]{Eq.~(\ref{e:#1})}
\newcommand{\EQ}[1]{(\ref{e:#1})}
\newcommand{\eqtwo}[2]{equations~(\ref{e:#1}) and~(\ref{e:#2})}
\newcommand{\EQTWO}[2]{Equations~(\ref{e:#1}) and~(\ref{e:#2})}
\newcommand{\fig}[1]{Fig.~\ref{f:#1}}
\newcommand{\FIG}[1]{Fig.~\ref{f:#1}}
\newcommand{\quot}[1]{\lq#1\rq}
\newcommand{\eg}{\textrm{e.g.}}
\newcommand{\cf}{\textrm{cf}}
\newcommand{\etc}{\textrm{etc}}
\newcommand{\ie}{\textrm{i.e.}}
\newcommand{\SET}[1]{\{#1\}}
\newcommand{\expl}[1]{\exp \left[ #1 \right] } 
\newcommand{\lb}{\left[}  
\newcommand{\rb}{\right]}  
\newcommand{\lc}{\left(}  
\newcommand{\rc}{\right)}  
\newcommand{\mult}{\times} 
\newcommand{\multcc}{\cdot} 
\newcommand{\multcn}{\cdot} 
\newcommand{\multnn}{\cdot} 
\newcommand{\ran}{\sub{ran}}
\newcommand{\dd}[1]{\text{d}{#1\ }}   
\newcommand{\ddd}[1]{\text{d}{#1}}   
\newcommand{\scal}[2]{(#1 \pmb{\cdot} #2)}
\newcommand{\mean}[1]{\overline{#1}}
\newcommand{\half}{\frac{1}{2}}
\newcommand{\Eesc}[1]{E^{\text{esc}}_{#1}}
\newcommand{\Lopt}{L_{\text{opt}}}
\newcommand{\dep}{\text{dep}}
\newcommand{\equil}{\text{eq}}
\newcommand{\Lrelax}{L_{\text{relax}}}
\newcommand{\therm}{\text{therm}}

\title{Dynamics below the depinning threshold} \author{Alejandro B.
Kolton}\email{kolton@physics.unige.ch} \affiliation{Universit\'e de Gen\`eve,
DPMC, 24 Quai Ernest Ansermet, CH-1211 Gen\`eve 4, Switzerland}
\author{Alberto Rosso} \email{rosso@lptms.u-psud.fr} \affiliation{CNRS;
Univ. Paris-Sud, UMR 8626, ORSAY CEDEX, F-91405, LPTMS}
\author{Thierry
Giamarchi}\email{giamarch@physics.unige.ch} \affiliation{Universit\'e
de Gen\`eve, DPMC, 24 Quai Ernest Ansermet, CH-1211 Gen\`eve 4,
Switzerland} \author{Werner Krauth} \email{krauth@lps.ens.fr}
\affiliation{CNRS-Laboratoire de Physique Statistique\\ Ecole Normale
Sup{\'{e}}rieure, 24 rue Lhomond, 75231 Paris Cedex 05, France}

\begin{abstract}
We study the steady-state low-temperature dynamics of an elastic
line in a disordered medium below the depinning threshold.
Analogously to the equilibrium dynamics, in the limit $T \rightarrow
0$, the steady state is dominated by a single configuration which is
occupied with probability one. We develop an exact algorithm to
target this dominant configuration and to analyze its geometrical
properties as a function of the driving force. The roughness
exponent of the line at large scales is identical to the one at
depinning. No length scale diverges in the steady state regime as
the  depinning threshold is approached from below. We do find,
a divergent length, but it is  associated only with the transient 
relaxation between metastable states.
\end{abstract}
\maketitle

The physics of elastic systems in disordered media has been
the focus of intense theoretical and experimental studies. A
continuing challenge has been to understand their response to an
applied external force $f$. Such a situation is relevant for a large
number of experimental driven systems ranging from periodic ones,
such as vortex lattices \cite{fuchs_creep_bglass} and charge density
waves \cite{lemay_creep_cdw} to interfaces, such as domain walls in
magnetic \cite{lemerle_domainwall_creep,repain_avalanches_magnetic} or
ferroelectric \cite{tybell_ferro_creep,paruch_2.5} materials, contact
lines of liquid menisci \cite{moulinet_distribution_width_contact_line}
and crack propagation \cite{ponson_fracture,maloy}.

A crucial feature of the zero temperature motion of these systems is
the existence of a threshold force $f_c$ below which the system is
pinned. For $f > f_c$ the system undergoes a depinning transition
\cite{narayan_fisher_cdw,nattermann_stepanow_depinning,chauve_zeta_twoloops,rosso_hartmann,duemmer}
and moves with a non-zero average velocity. Fisher first viewed the
depinning transition as a critical phenomenon
\cite{fisher_depinning_meanfield}. A key idea is that the slow
motion close to $f_c$ proceeds by avalanches of size $\xi$, that
diverge as $f_c$ is approached. Indeed, theoretical and numerical works
at zero temperature have shown that for $f \rightarrow f_c^+$, the
correlation length diverges as $\xi \sim (f-f_c)^{-\nu_{\dep}}$. The
exponent $\nu_{\dep}$ is related to the roughness exponent
$\zeta_{\dep}$ of the system pinned at $f_c$,  via the scaling
relation $\nu_{\dep}=1/(2-\zeta_{\dep})$. The length scale $\xi$
identifies both the typical size of avalanches and the crossover in
the roughness behavior \cite{duemmer}. For length scales below
$\xi$, the roughness is described by the exponent $\zeta_{\dep}$,
while for scales larger than $\xi$, the velocity introduces a time
dependent noise. In this regime, for interfaces, quenched disorder 
becomes equivalent to an effective temperature, and the roughness exponent is
equal to the thermal exponent. The interpretation of 
depinning as a critical phenomenon suggests that a similar diverging
length scale would also exists below $f_c$. Indeed, in standard
critical phenomena, for length scales smaller than the correlation
length the system is critical, and crosses over beyond this
correlation length to the broken symmetry phase on one side of the
transition and to the symmetric phase on the other side. An
important question is thus whether an equivalent correlation length
is observable in the limit $f \rightarrow f_c^-$.

Answering such a question is not an easy task since below $f_c$ the
steady-state velocity vanishes in the long-time limit at zero
temperature. This makes methods such as molecular dynamics
simulations ill suited. Some authors
\cite{middleton_depinning_below_fc,middleton_narayan,chen_marchetti} have therefore
studied the relaxation of a given initial configuration towards a
final zero-velocity state. This analysis allows to identify a
dynamical scaling characterized by a single correlation length
diverging with an exponent $\nu_{\dep}$ as $f_c$ is approached.
However it has remained unclear how such $T=0$ transients relate to the steady
state motion in the large time limit at small but finite
temperatures. Below $f_c$, it is thus preferable to keep the
temperature finite, but direct Langevin dynamics simulations are
powerless to clarify whether the limit $f \rightarrow f_c^-$ is also
characterized by a divergent length scale. An analytical tool for adressing
this issue is the Functional Renormalization Group
(FRG). This method allows to study, at finite temperature, the slow 
creep motion
\cite{ioffe_creep,nattermann_rfield_rbond,chauve_creep_long,kolton_string_creep}
taking place for $f \ll f_c$. In the creep regime, scaling
arguments rely on the physical properties of the system at
equilibrium ($f=0$). This phenomenological approach suggests that,
for $f \ll f_c$, the macroscopic forward motion is produced by
activated jumps with typical size $\xi_T \sim f^{-\nu_{\equil}}$.
The exponent $\nu_{\equil}$ can be related to the roughness by the
same scaling relation, $\nu_{\equil}=1/(2-\zeta_{\equil})$, but
with $\zeta_{\equil}$ the equilibrium roughness exponent. The
FRG analysis \cite{chauve_creep_long} shows that $\xi_T$ play the
role of crossover length between two roughness regimes: on length
scales below $\xi_T$, the roughness is described by the exponent
$\zeta_{\equil}$, while for scales  larger than $\xi_T$ it is described by the
exponent $\zeta_{\dep}$. This regime crosses over, for interfaces, to
a pure thermal-like behavior at very large length scales due to the
finiteness of the velocity for any non-zero temperature.  This FRG
finding is thus at variance with the interpretation of the
depinning transition as a standard critical phenomenon, since it
shows that the roughness $\zeta_{\dep}$ appears at large length
scales, and not short ones.

In order to address the question of the behavior below threshold, it
is interesting to consider the small-temperature limit without
however going all the way to $T=0$, so that steady-state motion
still exists. This limit is analogous to the $T\rightarrow 0$ limit
for the dynamics of a system at thermal equilibrium, where the
Boltzmann weights impose a ground state occupation with probability
one.  Occupation probabilities also exist for the steady-state
dynamics in a finite system, but they are more difficult to compute.
The $T\rightarrow 0$ limit of this dynamical regime is thus
described by a single dominant configuration, occupied with
probability one. This is particularly transparent for a particle
hopping on a one-dimensional ring
\cite{derrida_hopping_particle,ledoussal_creep_1d,ledoussal_private}.

\begin{figure}
\centerline{\includegraphics{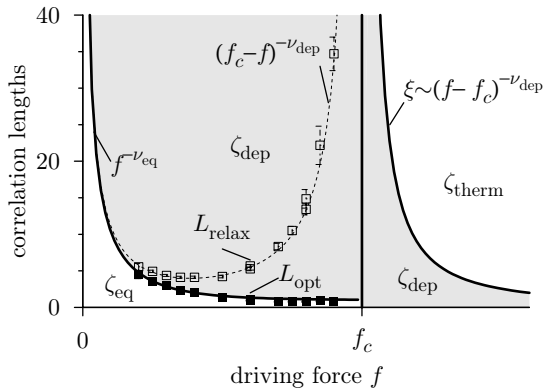}}
\caption{Dynamical phase diagram. The steady state 
properties of the elastic string at $T\rightarrow 0$ are
determined by $\Lopt$ (filled symbols) and $\xi$. They separate regions
characterized by the equilibrium exponent, the depinning exponent
(gray region), and the  thermal one. The divergent length $\Lrelax$
(open symbols) is associated only with transient dynamics. Lines are guides to the eye.}
\label{f:phasediagram}
\end{figure}

In this paper, we construct this dominant configurations using a
novel exact algorithm. We analyze its geometrical properties for
driving forces between the equilibrium and the depinning threshold.
We find, in agreement with the FRG scenario for $f \ll f_c$, that
the depinning roughness exponent describes the geometry of the line
at large length scales for all forces. This result is inconsistent
with the existence of a divergent correlation length for $f
\rightarrow f_c^-$ \emph{in the steady state regime}.  A divergent
length scale exists only in the transient dynamics of relaxation
between two metastable states. The conclusions of our analysis are summarized in 
the dynamical phase diagram of \fig{phasediagram}.

\begin{figure} \centerline{\includegraphics{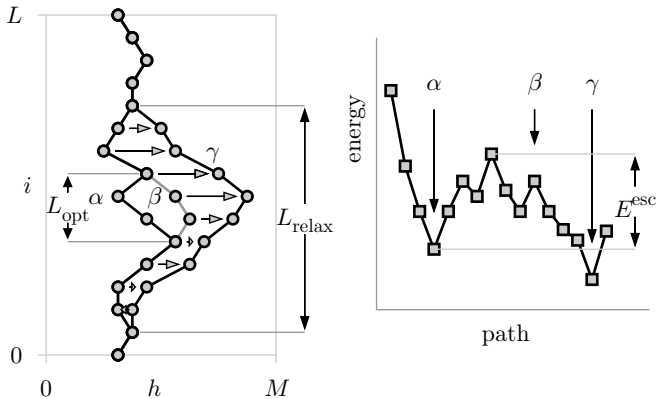}}
\caption{Low temperature dynamics of the driven elastic string below
the depinning threshold: the optimal path to escape from a given
metastable configuration $\alpha$ pass through a saddle
configuration $\beta$ that can relax deterministically to the next
metastable configuration $\gamma$ with $E_{\gamma} <E_{\alpha}$.
Configurations $\alpha $ and $\beta$ differ on a length $\Lopt$;
$\alpha$ and $\gamma$ differ on a length $\Lrelax$.}
\label{f:model}
\end{figure}

Let us consider an elastic line on a two-dimensional discrete $L\times
M$ lattice with periodic boundary conditions in both directions (see
\fig{model}). The force $f$ drives the line around the system in the
direction of $M$. We define elementary moves as in
Ref.~\cite{rosso_vmc_string,rosso_dep_exponent}, allowing for
simultaneous motion of several sites. This avoids problems of
single-site dynamics proper to the elastic string model. Using this
dynamics, any configuration relaxes to the nearest metastable
configuration, a local minimum of the energy
\begin{equation}
E = \sum_i  \frac{1}{2} (h_{i+1}- h_{i})^2  - f h_i + V(i,h_i).
\label{}
\end{equation}
The first term in this equation takes into account the short-range
elastic energy and $V(i,h_i)$ is the quenched disorder, which we
take to be Gaussian and uncorrelated.  The variables $h$ gives the
height of the line, as a function of $i$ (see \fig{model}).  The
equilibrium ground state (at $f=0$) is easily found with a transfer
matrix approach \cite{huse_exponent_line, kardar_exponent_line}. The
sample-dependent critical force $f_c$ and the zero-temperature
critical configuration at $f_c$ are also readily determined
\cite{rosso_vmc_string,rosso_dep_exponent} exploiting the analytic
structure of the zero-temperature dynamics of a $d$-dimensional
elastic interface embedded in a $d+1$ random medium. Two properties
are particularly useful for numerical algorithms: the 
\quot{no passing} property assures that, during its motion, an
elastic interface can never miss any pinned configuration; the
\quot{no return} property,  which holds after a finite initial
time, states that only forward
motion takes place \cite{middleton_theorem}. At non-zero
temperatures, these rules obviously cease to hold. The
motion is dominated by the activation time spent to overcome
the energy barriers. The evaluation of these barriers is an
NP-complete problem \cite{middleton_NP_complete} and the algorithms
employed at the equilibrium are exponential in the size of the
system.  In the following we show that, in the $T \rightarrow 0$
limit, properties analogous to the no-passing rule govern the time
evolution of such systems. These properties allow to capture the
steady state dynamics below the depinning threshold within a reasonable
computation time.

For an elastic line pinned in a metastable configuration $\alpha$,
the path taken at low temperature from $\alpha$ to another
metastable configuration with lower energy $\gamma$ is the one with
the lowest activation energy $\Eesc{\alpha}$. The time spent in
$\alpha$ thus corresponds to the Arrhenius activation energy
$\Eesc{\alpha}$. Once at $\gamma$, the probability to return to
$\alpha$ are neglegible if $T \rightarrow 0$. This procedure
defines a deterministic coarse-grained dynamics between metastable
states. Two theorems can be proved for this dynamics, whenever the
standard no-passing rule is valid: (i) If there is no configuration
which lowers the energy of $\alpha$ in the backward direction, the
coarse-grained dynamics starting from $\alpha$ is always
forward-directed; (ii) Let $\alpha$ be any metastable configuration
escaping into a configuration $\gamma$ with $h^{\gamma} \ge
h^{\alpha}$ and $\gamma'$ any configuration such that $h^{\gamma'}
\ge h^{\alpha}$ and having an energy barrier $\Eesc{\gamma'} >
\Eesc{\alpha}$: all $\gamma'$ then satisfy $h^{\gamma'} \ge
h^{\gamma}$ \cite{longpaper}. Direct consequences of these theorems
are that the dynamics is periodic after a single pass of the line
around the system, and that the dominant configuration, the metastable configuration with the
largest barrier,  is always visited, independently of the initial
condition. This dominant metastable configuration is the only
statistically relevant configuration of the $T \rightarrow 0$ steady
state dynamics: Under the conditions of Arrhenius activation,  the
system spends much more time in it than in any other
configuration. To construct the minimal path escaping from the
configuration $\alpha$ (see \fig{model}),
we enumerate all configurations dynamically connected to it,
increasing gradually the maximum energy of configurations.  This
continues until a saddle configuration $\beta$ is found which can
relax to a metastable configuration $\gamma$, with
$E_{\gamma}<E_{\alpha}$. Configuration $\alpha $ and $\beta$ differ on
a length $\Lopt$; $\alpha $ and $\gamma$ differ on a length $\Lrelax$.
A new path construction starts at configuration $\gamma$.  Several
refinements are implemented \cite{longpaper}.  During the
construction, a very large number of configurations are considered,
but only those with energy barriers below $\Eesc{\alpha}$ need to be
evolved. In our case, the computational cost increases very rapidly
with $\Lopt$, but small values of $\Lopt$ can be handled even for
large system sizes $L$. This is realized in the physically interesting
case close to the depinning threshold and the computation becomes
difficult only for small driving forces, where the saddle point
configuration may be very different from $\alpha$.  We have considered
systems up to $L=128$ for forces $f \gtrsim 0.7 f_c$ and up to $L =
32$ for $f \sim 0.2 f_c$.

\begin{figure} \centerline{ \includegraphics{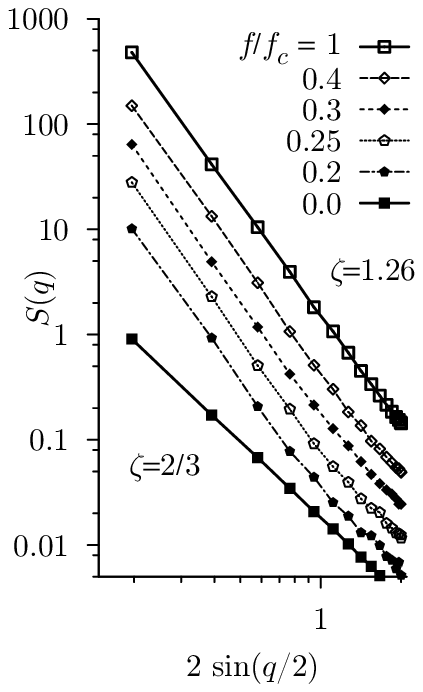}
\includegraphics{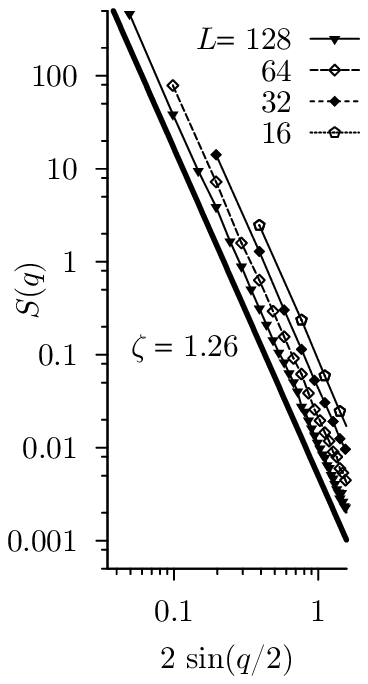}} \caption{Steady state
structure factor of the line in the $T \rightarrow 0$ limit, averaged
on $1000$ samples. (a) $S(q)$ for $L=32$, $M=92$ and different forces
(curves are shifted for clarity). (b) $S(q)$ at $f/f_c=0.8$
for $L=16,32,64,128$ and $M=L^{\zeta_{\dep}}$.    }
\label{f:structure}
\end{figure}
In the following, we concentrate on the disorder-averaged geometric
properties of the dominant configuration, expressed through the structure
factor $S(q)$, which gives access to the roughness exponent $\zeta$:
\begin{equation} S(q)= \mean{ | \sum_j h_j \exp(- i q j) |^2} \sim q^{-(1
+ 2 \zeta) } \label{} \end{equation} We also analyze the dynamical
lengths $\Lopt$ and $\Lrelax$ defined above. In \fig{structure}(a),
we show $S(q)$ for different forces ranging from $f=0$ (equilibrium)
to $f=f_c$ (depinning). The slope of $S(q)$ in these two limiting
cases corresponds to the critical exponents $\zeta_{\equil}= 2/3$
\cite{kardar_zeta_dp} and $\zeta_{\dep}=1.26 \pm 0.01$
\cite{rosso_hartmann}, respectively. In between, we find two
regimes. Remarkably, on large length scales (small $q$), $S(q)$
corresponds to $\zeta_{\dep}$, even at forces well below $f_c$. In
\fig{structure}(b) we show that this behavior is size independent up
to $L=128$, at $f=0.8 f_c$. Only at small scales (large $q$) the
roughness exponent is $\zeta_{\equil}$. This allows to define a
crossover length which decreases with increasing $f$, as we can
observe in \fig{structure}(a). The results of \fig{structure} thus
show that there is no divergent length scale below the depinning
threshold in the {\it steady state} regime, unlike the divergent
length scale that exists above the depinning threshold.

In order to understand the existence of a single characteristic length
which does not diverge at $f_c$ but separates the $\zeta_{\equil}$ and
$\zeta_{\dep}$ steady state regimes, we analyze the behavior of the
dynamical lengths $\Lopt$ and $\Lrelax$ as a function of $f$. The
dynamical lengths shown in \fig{phasediagram} are obtained starting
from the dominant configuration and averaged on $1000$ samples. We see
that $\Lopt$ decreases with increasing $f$ and saturates to the
minimal avalanche size at $f_c$. Interestingly, $\Lrelax$ increases as
$(f_c-f)^{-\nu_{\dep}}$ close to $f_c$ and $\Lrelax \approx \Lopt$ at
low $f$. This behavior is controlled by the density of metastable
states, which is very high at low forces, and very low near $f_c$. The
results of \fig{phasediagram} lead us thus to identify $\Lopt$ with
the steady state crossover length. The observed divergence in
$\Lrelax$ has its origin purely in transient dynamics and has no
counterpart in the steady state properties. We note that $\Lrelax$ can
be related with the divergent length found in
\cite{middleton_depinning_below_fc,middleton_narayan,chen_marchetti}. Our
simulations allow to sketch the dynamical phase diagram of
\fig{phasediagram}. This diagram is expected to hold for
$d$-dimensional manifolds in $d+1$ dimensions. The steady state
geometrical properties of the string in the $T\rightarrow 0$ limit are
determined by the two solid lines in \fig{phasediagram}: $\Lopt$,
associated with the optimal activated jump needed to escape from the
dominant configuration, and $\xi$, the typical size of the depinning
avalanche above threshold. $\Lopt(f)$ and $\xi(f)$ separate three
roughness regimes: $\zeta_{\equil}=2/3$, $\zeta_{\dep}=1.26$ and
$\zeta_{\therm}=0.5$. The divergent length $\Lrelax$ (dashed line) does
not affect steady state properties: it describes transient processes
that depend on the distance between successive metastable states, but
it does not describe the properties of the dominant ones. As indicated
in \fig{phasediagram}, it is plausible to connect continuously $\Lopt$
with the thermal nucleus size $\xi_T$ defined for $f \ll
f_c$. However, we are not able to verify with our data the predicted
power law divergence $\xi_T \sim f^{-\nu_{\equil}}$. The scenario thus
proposed is consistent with the FRG prediction
\cite{chauve_creep_long} of the existence of $\zeta_{\dep}$ at scales
larger than $\xi_T$ and inconsistent with the existence of a critical
region extending both below and above $f_c$, as the standard
critical phenomena interpretation would have suggested.

Although our computational approach strictly holds only in the $T \rightarrow
0$ limit, it yields the scenario of \fig{phasediagram}, which offers a solid
framework for discussing finite temperature effects. At finite temperature the
depinning transition is rounded. Since the velocity is finite for all forces,
the regime with the thermal roughness $\zeta_{\therm}$ exists at the largest
length scales \cite{chauve_creep_long}. For $f<f_c$ we expect two crossovers
separating the regimes $\zeta_{\equil}$, $\zeta_{\dep}$ and $\zeta_{\therm}$,
as we increase the length scale. For temperature comparable to the strength of
the disorder at the very small length scales an additional thermal regime
appears and crosses over to $\zeta_{\equil}$ or directly to
$\zeta_{\dep}$. Such a scenario is in part supported by recent Langevin
dynamics simulations \cite{kolton_string_creep}, which show, at small forces,
a steady state motion characterized by a roughness exponent bigger than
$\zeta_{\equil}$ at larges length scales and a roughness $\zeta_{\therm}$ at
the smallest length scales.

Finally, we expect that an experimental verification of our results is
possible using, for instance, imaging techniques for magnetic
\cite{lemerle_domainwall_creep,repain_avalanches_magnetic} or electric
\cite{tybell_ferro_creep,paruch_2.5} domain walls in thin films:
$\Lopt$ could be extracted from the analysis of a spatial correlation
function or by measuring the domain wall speed at low temperatures,
since $\Lopt$ also controls the thermally activated motion
\cite{longpaper}; $\Lrelax$ could be measured by comparing consecutive
(long-lived) metastable states when $f$ is close to $f_c$, since then
$\Lrelax$ controls the distances between successive metastable states.


We thank P.~Le~Doussal for very helpful discussions all along this
work and also acknowledge stimulating discussions with C.~J.~Bolech
and A.~A.~Middleton. We are grateful to J.~Albiero for useful
programming advice. A.~R. and  W.~K. thank DPMC in Geneva for hospitality during
part of this work. This work was supported in part by the Swiss
National Fund under MANEP and division II.



\begin{thebibliography}{32}
\expandafter\ifx\csname natexlab\endcsname\relax\def\natexlab#1{#1}\fi
\expandafter\ifx\csname bibnamefont\endcsname\relax
  \def\bibnamefont#1{#1}\fi
\expandafter\ifx\csname bibfnamefont\endcsname\relax
  \def\bibfnamefont#1{#1}\fi
\expandafter\ifx\csname citenamefont\endcsname\relax
  \def\citenamefont#1{#1}\fi
\expandafter\ifx\csname url\endcsname\relax
  \def\url#1{\texttt{#1}}\fi
\expandafter\ifx\csname urlprefix\endcsname\relax\def\urlprefix{URL }\fi
\providecommand{\bibinfo}[2]{#2}
\providecommand{\eprint}[2][]{\url{#2}}

\bibitem[{\citenamefont{{Fuchs {\it et al.}}}(1998)}]{fuchs_creep_bglass}
\bibinfo{author}{\bibfnamefont{D.~T.} \bibnamefont{{Fuchs {\it et al.}}}},
  \bibinfo{journal}{Phys. Rev. Lett.} \textbf{\bibinfo{volume}{80}},
  \bibinfo{pages}{4971} (\bibinfo{year}{1998}).

\bibitem[{\citenamefont{Lemay et~al.}(1999)\citenamefont{Lemay, Thorne, Li, and
  Brock}}]{lemay_creep_cdw}
\bibinfo{author}{\bibfnamefont{S.~G.} \bibnamefont{Lemay}},
  \bibinfo{author}{\bibfnamefont{R.~E.} \bibnamefont{Thorne}},
  \bibinfo{author}{\bibfnamefont{Y.}~\bibnamefont{Li}}, \bibnamefont{and}
  \bibinfo{author}{\bibfnamefont{J.~D.} \bibnamefont{Brock}},
  \bibinfo{journal}{Phys. Rev. Lett.} \textbf{\bibinfo{volume}{83}},
  \bibinfo{pages}{2793} (\bibinfo{year}{1999}).

\bibitem[{\citenamefont{Lemerle et~al.}(1998)\citenamefont{Lemerle, Ferr{\'e},
  Chappert, Mathet, Giamarchi, and {Le Doussal}}}]{lemerle_domainwall_creep}
\bibinfo{author}{\bibfnamefont{S.}~\bibnamefont{{Lemerle {\it et al.}}}},
  \bibinfo{journal}{Phys. Rev. Lett.}
  \textbf{\bibinfo{volume}{80}}, \bibinfo{pages}{849} (\bibinfo{year}{1998}).

\bibitem[{\citenamefont{Repain et~al.}(2004)\citenamefont{Repain, Bauer, Jamet,
  Ferr{\'e}, Mougin, Chappert, and Bernas}}]{repain_avalanches_magnetic}
\bibinfo{author}{\bibfnamefont{V.}~\bibnamefont{{Repain {\it et al.}}}},
  \bibinfo{journal}{Europhys. Lett.} \textbf{\bibinfo{volume}{68}},
  \bibinfo{pages}{460} (\bibinfo{year}{2004}).

\bibitem[{\citenamefont{Tybell et~al.}(2002)\citenamefont{Tybell, Paruch,
  Giamarchi, and Triscone}}]{tybell_ferro_creep}
\bibinfo{author}{\bibfnamefont{T.}~\bibnamefont{Tybell}},
  \bibinfo{author}{\bibfnamefont{P.}~\bibnamefont{Paruch}},
  \bibinfo{author}{\bibfnamefont{T.}~\bibnamefont{Giamarchi}},
  \bibnamefont{and} \bibinfo{author}{\bibfnamefont{J.~M.}
  \bibnamefont{Triscone}}, \bibinfo{journal}{Phys. Rev. Lett.}
  \textbf{\bibinfo{volume}{89}}, \bibinfo{pages}{097601}
  (\bibinfo{year}{2002}).

\bibitem[{\citenamefont{Paruch et~al.}(2005)\citenamefont{Paruch, Triscone, and
  Giamarchi}}]{paruch_2.5}
\bibinfo{author}{\bibfnamefont{P.}~\bibnamefont{Paruch}},
  \bibinfo{author}{\bibfnamefont{J.~M.} \bibnamefont{Triscone}},
  \bibnamefont{and}
  \bibinfo{author}{\bibfnamefont{T.}~\bibnamefont{Giamarchi}},
  \bibinfo{journal}{Phys. Rev. Lett.} \textbf{\bibinfo{volume}{94}},
  \bibinfo{pages}{197601} (\bibinfo{year}{2005}).

\bibitem[{\citenamefont{Moulinet et~al.}(2004)\citenamefont{Moulinet, Rosso,
  Krauth, and Rolley}}]{moulinet_distribution_width_contact_line}
\bibinfo{author}{\bibfnamefont{S.}~\bibnamefont{Moulinet}},
  \bibinfo{author}{\bibfnamefont{A.}~\bibnamefont{Rosso}},
  \bibinfo{author}{\bibfnamefont{W.}~\bibnamefont{Krauth}}, \bibnamefont{and}
  \bibinfo{author}{\bibfnamefont{E.}~\bibnamefont{Rolley}},
  \bibinfo{journal}{Phys. Rev. E} \textbf{\bibinfo{volume}{69}},
  \bibinfo{pages}{035103} (\bibinfo{year}{2004}).

\bibitem[{\citenamefont{Ponson et~al.}(2006)\citenamefont{Ponson, Bonamy, and
  Bouchaud}}]{ponson_fracture}
\bibinfo{author}{\bibfnamefont{L.}~\bibnamefont{Ponson}},
  \bibinfo{author}{\bibfnamefont{D.}~\bibnamefont{Bonamy}}, \bibnamefont{and}
  \bibinfo{author}{\bibfnamefont{E.}~\bibnamefont{Bouchaud}},
  \bibinfo{journal}{Phys. Rev. Lett.} \textbf{\bibinfo{volume}{96}},
  \bibinfo{pages}{035506} (\bibinfo{year}{2006}).

\bibitem[{\citenamefont{M{\aa}l{\o}y and Schmittbuhl}(2001)}]{maloy}
\bibinfo{author}{\bibfnamefont{K.~J.} \bibnamefont{M{\aa}l{\o}y}}
  \bibnamefont{and}
  \bibinfo{author}{\bibfnamefont{J.}~\bibnamefont{Schmittbuhl}},
  \bibinfo{journal}{Phys. Rev. Lett.} \textbf{\bibinfo{volume}{87}},
  \bibinfo{pages}{105502} (\bibinfo{year}{2001}).

\bibitem[{\citenamefont{Narayan and Fisher}(1992)}]{narayan_fisher_cdw}
\bibinfo{author}{\bibfnamefont{O.}~\bibnamefont{Narayan}} \bibnamefont{and}
  \bibinfo{author}{\bibfnamefont{D.~S.} \bibnamefont{Fisher}},
  \bibinfo{journal}{Phys. Rev. B} \textbf{\bibinfo{volume}{46}},
  \bibinfo{pages}{11520} (\bibinfo{year}{1992}).

\bibitem[{\citenamefont{Nattermann et~al.}(1992)\citenamefont{Nattermann,
  Stepanow, Tang, and Leschhorn}}]{nattermann_stepanow_depinning}
\bibinfo{author}{\bibfnamefont{T.}~\bibnamefont{Nattermann}},
  \bibinfo{author}{\bibfnamefont{S.}~\bibnamefont{Stepanow}},
  \bibinfo{author}{\bibfnamefont{L.~H.} \bibnamefont{Tang}}, \bibnamefont{and}
  \bibinfo{author}{\bibfnamefont{H.}~\bibnamefont{Leschhorn}},
  \bibinfo{journal}{J. Phys. (Paris)} \textbf{\bibinfo{volume}{2}},
  \bibinfo{pages}{1483} (\bibinfo{year}{1992}).

\bibitem[{\citenamefont{Chauve et~al.}(2001)\citenamefont{Chauve, Wiese, and
  {Le Doussal}}}]{chauve_zeta_twoloops}
\bibinfo{author}{\bibfnamefont{P.}~\bibnamefont{Chauve}},
  \bibinfo{author}{\bibfnamefont{K.}~\bibnamefont{Wiese}}, \bibnamefont{and}
  \bibinfo{author}{\bibfnamefont{P.}~\bibnamefont{{Le Doussal}}},
  \bibinfo{journal}{Phys. Rev. Lett.} \textbf{\bibinfo{volume}{86}},
  \bibinfo{pages}{1785} (\bibinfo{year}{2001}).

\bibitem[{\citenamefont{Rosso et~al.}(2003)\citenamefont{Rosso, Hartmann, and
  Krauth}}]{rosso_hartmann}
\bibinfo{author}{\bibfnamefont{A.}~\bibnamefont{Rosso}},
  \bibinfo{author}{\bibfnamefont{A.~K.} \bibnamefont{Hartmann}},
  \bibnamefont{and} \bibinfo{author}{\bibfnamefont{W.}~\bibnamefont{Krauth}},
  \bibinfo{journal}{Phys. Rev. E} \textbf{\bibinfo{volume}{67}},
  \bibinfo{pages}{021602} (\bibinfo{year}{2003}).

\bibitem[{\citenamefont{Duemmer and Krauth}(2005)}]{duemmer}
\bibinfo{author}{\bibfnamefont{O.}~\bibnamefont{Duemmer}} \bibnamefont{and}
  \bibinfo{author}{\bibfnamefont{W.}~\bibnamefont{Krauth}},
  \bibinfo{journal}{Phys. Rev. E} \textbf{\bibinfo{volume}{71}},
  \bibinfo{pages}{061601} (\bibinfo{year}{2005}).

\bibitem[{\citenamefont{Fisher}(1985)}]{fisher_depinning_meanfield}
\bibinfo{author}{\bibfnamefont{D.~S.} \bibnamefont{Fisher}},
  \bibinfo{journal}{Phys. Rev. B} \textbf{\bibinfo{volume}{31}},
  \bibinfo{pages}{1396} (\bibinfo{year}{1985}).

\bibitem[{\citenamefont{Middleton and
  Fisher}(1993)}]{middleton_depinning_below_fc}
\bibinfo{author}{\bibfnamefont{A.~A.} \bibnamefont{Middleton}}
  \bibnamefont{and} \bibinfo{author}{\bibfnamefont{D.~S.}
  \bibnamefont{Fisher}}, \bibinfo{journal}{Phys. Rev. B}
  \textbf{\bibinfo{volume}{47}}, \bibinfo{pages}{3530} (\bibinfo{year}{1993}).

\bibitem[{\citenamefont{Narayan and Middleton}(1994)}]{middleton_narayan}
\bibinfo{author}{\bibfnamefont{O.} \bibnamefont{Narayan}}
  \bibnamefont{and} \bibinfo{author}{\bibfnamefont{A.~A.}
  \bibnamefont{Middleton}}, \bibinfo{journal}{Phys. Rev. B}
  \textbf{\bibinfo{volume}{49}}, \bibinfo{pages}{244} (\bibinfo{year}{1994}).


\bibitem[{\citenamefont{Chen and Marchetti}(1995)}]{chen_marchetti}
\bibinfo{author}{\bibfnamefont{L.~W.} \bibnamefont{Chen}} \bibnamefont{and}
  \bibinfo{author}{\bibfnamefont{M.~C.} \bibnamefont{Marchetti}},
  \bibinfo{journal}{Phys. Rev. B} \textbf{\bibinfo{volume}{51}},
  \bibinfo{pages}{6296} (\bibinfo{year}{1995}).

\bibitem[{\citenamefont{Ioffe and Vinokur}(1987)}]{ioffe_creep}
\bibinfo{author}{\bibfnamefont{L.~B.} \bibnamefont{Ioffe}} \bibnamefont{and}
  \bibinfo{author}{\bibfnamefont{V.~M.} \bibnamefont{Vinokur}},
  \bibinfo{journal}{J. Phys. C} \textbf{\bibinfo{volume}{20}},
  \bibinfo{pages}{6149} (\bibinfo{year}{1987}).

\bibitem[{\citenamefont{Nattermann}(1987)}]{nattermann_rfield_rbond}
\bibinfo{author}{\bibfnamefont{T.}~\bibnamefont{Nattermann}},
  \bibinfo{journal}{Europhys. Lett.} \textbf{\bibinfo{volume}{4}},
  \bibinfo{pages}{1241} (\bibinfo{year}{1987}).

\bibitem[{\citenamefont{Chauve et~al.}(2000)\citenamefont{Chauve, Giamarchi,
  and {Le Doussal}}}]{chauve_creep_long}
\bibinfo{author}{\bibfnamefont{P.}~\bibnamefont{Chauve}},
  \bibinfo{author}{\bibfnamefont{T.}~\bibnamefont{Giamarchi}},
  \bibnamefont{and} \bibinfo{author}{\bibfnamefont{P.}~\bibnamefont{{Le
  Doussal}}}, \bibinfo{journal}{Phys. Rev. B} \textbf{\bibinfo{volume}{62}},
  \bibinfo{pages}{6241} (\bibinfo{year}{2000}).

\bibitem[{\citenamefont{Kolton et~al.}(2005)\citenamefont{Kolton, Rosso, and
  Giamarchi}}]{kolton_string_creep}
\bibinfo{author}{\bibfnamefont{A.~B.} \bibnamefont{Kolton}},
  \bibinfo{author}{\bibfnamefont{A.}~\bibnamefont{Rosso}}, \bibnamefont{and}
  \bibinfo{author}{\bibfnamefont{T.}~\bibnamefont{Giamarchi}},
  \bibinfo{journal}{Phys. Rev. Lett.} \textbf{\bibinfo{volume}{94}},
  \bibinfo{pages}{047002} (\bibinfo{year}{2005}).

\bibitem[{\citenamefont{Derrida}(1983)}]{derrida_hopping_particle}
\bibinfo{author}{\bibfnamefont{B.}~\bibnamefont{Derrida}}, \bibinfo{journal}{J.
  Stat. Phys.} \textbf{\bibinfo{volume}{31}}, \bibinfo{pages}{433}
  (\bibinfo{year}{1983}).

\bibitem[{\citenamefont{{Le Doussal} and Vinokur}(1995)}]{ledoussal_creep_1d}
\bibinfo{author}{\bibfnamefont{P.}~\bibnamefont{{Le Doussal}}}
  \bibnamefont{and} \bibinfo{author}{\bibfnamefont{V.~M.}
  \bibnamefont{Vinokur}}, \bibinfo{journal}{Physica C}
  \textbf{\bibinfo{volume}{254}}, \bibinfo{pages}{63} (\bibinfo{year}{1995}).

\bibitem[{\citenamefont{Doussal}()}]{ledoussal_private}
\bibinfo{author}{\bibfnamefont{P.} \bibnamefont{Le Doussal}},
  \bibinfo{note}{private communication}.

\bibitem[{\citenamefont{Rosso and Krauth}(2002)}]{rosso_vmc_string}
\bibinfo{author}{\bibfnamefont{A.}~\bibnamefont{Rosso}} \bibnamefont{and}
  \bibinfo{author}{\bibfnamefont{W.}~\bibnamefont{Krauth}},
  \bibinfo{journal}{Phys. Rev. B} \textbf{\bibinfo{volume}{65}},
  \bibinfo{pages}{012202} (\bibinfo{year}{2002}).

\bibitem[{\citenamefont{Rosso and Krauth}(2001)}]{rosso_dep_exponent}
\bibinfo{author}{\bibfnamefont{A.}~\bibnamefont{Rosso}} \bibnamefont{and}
  \bibinfo{author}{\bibfnamefont{W.}~\bibnamefont{Krauth}},
  \bibinfo{journal}{Phys. Rev. Lett.} \textbf{\bibinfo{volume}{87}},
  \bibinfo{pages}{187002} (\bibinfo{year}{2001}).

\bibitem[{\citenamefont{Huse and Henley}(1985)}]{huse_exponent_line}
\bibinfo{author}{\bibfnamefont{D.~A.} \bibnamefont{Huse}} \bibnamefont{and}
  \bibinfo{author}{\bibfnamefont{C.~L.} \bibnamefont{Henley}},
  \bibinfo{journal}{Phys. Rev. Lett.} \textbf{\bibinfo{volume}{54}},
  \bibinfo{pages}{2708} (\bibinfo{year}{1985}).

\bibitem[{\citenamefont{Kardar}(1985)}]{kardar_exponent_line}
\bibinfo{author}{\bibfnamefont{M.}~\bibnamefont{Kardar}},
  \bibinfo{journal}{Phys. Rev. Lett.} \textbf{\bibinfo{volume}{55}},
  \bibinfo{pages}{2923} (\bibinfo{year}{1985}).

\bibitem[{\citenamefont{Middleton}(1992)}]{middleton_theorem}
\bibinfo{author}{\bibfnamefont{A.~A.} \bibnamefont{Middleton}},
  \bibinfo{journal}{Phys. Rev. Lett.} \textbf{\bibinfo{volume}{68}},
  \bibinfo{pages}{671} (\bibinfo{year}{1992}).

\bibitem[{\citenamefont{Middleton}(1999)}]{middleton_NP_complete}
\bibinfo{author}{\bibfnamefont{A.~A.} \bibnamefont{Middleton}},
  \bibinfo{journal}{Phys. Rev. E} \textbf{\bibinfo{volume}{59}},
  \bibinfo{pages}{2571} (\bibinfo{year}{1999}).

\bibitem[{\citenamefont{Kolton et~al.}()\citenamefont{Kolton, Rosso, Giamarchi,
  and Krauth}}]{longpaper}
\bibinfo{author}{\bibfnamefont{A.~B.} \bibnamefont{Kolton}},
  \bibinfo{author}{\bibfnamefont{A.}~\bibnamefont{Rosso}},
  \bibinfo{author}{\bibfnamefont{T.}~\bibnamefont{Giamarchi}},
  \bibnamefont{and} \bibinfo{author}{\bibfnamefont{W.}~\bibnamefont{Krauth}},
  \bibinfo{note}{to be published}.

\bibitem[{\citenamefont{Kardar}(1987)}]{kardar_zeta_dp}
\bibinfo{author}{\bibfnamefont{M.}~\bibnamefont{Kardar}},
  \bibinfo{journal}{Nucl. Phys. B} \textbf{\bibinfo{volume}{290}},
  \bibinfo{pages}{582} (\bibinfo{year}{1987}).

\end{thebibliography}


\end{document}